\documentclass[12pt,fleqn,a4paper,nofootinbib,aps,showkeys,prd,showpacs]{revtex4}
\usepackage[cp1250]{inputenc}
\usepackage[english]{babel}
\usepackage{amsmath,amssymb}
\allowdisplaybreaks
\newcommand{\journal}[1]{\textit{#1}}
\begin{document}

\title{Propagator of the interacting Rarita-Schwinger field}
\author{A.~E.~Kaloshin}
\email{kaloshin@physdep.isu.ru}
\author{V.~P.~Lomov}
\email{lomov@physdep.isu.ru}
\affiliation{Irkutsk State University, K.~Marks Str. 1, 664003
Irkutsk, Russia.}
\begin{abstract}
We obtain in analytical form the dressed propagator of the massive
Rarita-Schwinger field taking into account all spin components and
discuss shortly its properties.
\end{abstract}
\keywords{Rarita-Schwinger field, dressed propagator}
\pacs{11.10.-z, 11.20.Gx} \maketitle

\section*{Introduction}

The covariant description of the spin 3/2 particles is usually
based on the Rarita-Schwinger formalism \cite{Rar-Sch}
where the main object is the spin-vector field $\Psi^{\mu}$.
However in addition to spin 3/2 this field contains extra spin 1/2
components and this circumstance generates the main difficulties
in its description \cite{JohnSud61,VeZw69}. The problem has a long history, we mention
here only relatively recent works \cite{Kor,Pas98,Pas01,Cas,Kir,AKS} (see
for older references therein) which contain some discussion of the
problem.

The most general lagrangian for free Rarita-Schwinger field has
the following form(see e.g. \cite{AU,NEK,BDM}):
\begin{flalign}
\mathcal{L}=&\overline{\Psi}
\vphantom{\Psi}^{\mu}\Lambda^{\mu\nu}\Psi^{\nu},\notag\\
\Lambda^{\mu\nu}=&(\hat{p}-M)g^{\mu\nu}+A(\gamma^{\mu}p^{\nu}+
\gamma^{\nu}p^{\mu})+\frac{1}{2}(3A^2+2A+1)\gamma^{\mu}
\hat{p}\gamma^{\nu}+M(3A^2+3A+1)\gamma^{\mu}\gamma^{\nu}.
\label{lagr}
\end{flalign}
Here $M$ is the mass of Rarita-Schwinger field, $A$ is an arbitrary
parameter, $p_{\mu}=i \partial_{\mu}$.

This lagrangian is invariant under the point transformation:
\begin{flalign*}
\Psi^{\mu}\to\Psi^{\prime\,\mu}=(g^{\mu\nu}+\alpha\gamma^{\mu}\gamma^{\nu})\Psi^{\nu},
\ \ \ \ \ \ \ \ A\to A^{\prime}=\frac{A-2\alpha}{1+4\alpha},
\end{flalign*}
with parameter $\alpha\neq -1/4$.

The lagrangian (\ref{lagr}) leads to the following equations of motion:
\begin{equation}
\Lambda^{\mu\nu}\Psi^{\nu}=0.
\end{equation}
The free propagator of the Rarita-Schwinger field
in a momentum space obeys the equation:
\begin{equation}
\Lambda^{\mu\nu}G_{0}\mkern -4mu {\vphantom{\Lambda}}^{\nu\rho}=g^{\mu\rho}.
\end{equation}
The expression for the free propagator $G^{\mu\nu}_0$ is well
known (see references in above), thus we do not present it here.

As concerned for the dressed propagator, its construction is a
more complicated issue and its total expression is unknown up to
now. Thus a practical use of $G^{\mu\nu}$ (in particular for the
case of $\Delta(1232)$ production) needs some approximation in its
description. The standard approximation \cite{Pas95,KS} consist in
a dressing of the spin 3/2 components only while the rest
components of $G^{\mu\nu}$ can be neglected or considered as bare.
Another way to take into account the spin 1/2 components is a
numerical solution of the appearing system of equations\cite{Kor,AKS}.
In Ref.~\cite{Pas95} it was noticed that the spin
1/2 components are necessary to reproduce the experimental data on
the $\Delta(1232)$ production in Compton scattering. So the
correct account of the extra spin 1/2 component in the
$\Psi^{\mu}$ field has also a practical meaning.

In this paper we derive an analytical expression for the
interacting Rarita-Schwinger field's propagator with accounting
all spin components and shortly discuss its properties. It turned
out that the spin 1/2 part of the dressed propagator has rather
compact form, and a crucial point for its deriving  is the
choosing of a suitable basis.

\section*{Dyson-Schwinger equation and its solution}

The Dyson-Schwinger equation for the propagator of the
Rarita-Schwinger field has the following form
\begin{equation}
G^{\mu\nu}=G^{\mu\nu}_{0}+G^{\mu\alpha}J^{\alpha\beta}G^{\beta\nu}_{0}.
\end{equation}
Here $G_{0}^{\mu\nu}$ and $G^{\mu\nu}$ are the free and full
propagators respectively, $J^{\mu\nu}$ is a self-energy
contribution. The equation may be rewritten for inverse
propagators as
\begin{equation}
(G^{-1})^{\mu\nu}=(G^{-1}_{0})^{\mu\nu}-J^{\mu\nu}.\label{inverD}
\end{equation}
If we consider the self-energy $J^{\mu\nu}$ as a known value
\footnote{That is the widely used in the resonance physics
"rainbow approximation", see \textit{e.g.} recent review\cite{MaRo}},
than the problem is reduced to reversing of relation
\eqref{inverD}. This procedure is not technically evident and it
needs some preparations. First of all it is useful to have a basis
for both propagators and self-energy.
\begin{enumerate}
\item The most natural basis for the spin-tensor $S^{\mu\nu}(p)$
decomposition is the $\gamma$-matrix one:
      \begin{equation}
      \begin{split}
      S^{\mu\nu}(p)=&g^{\mu\nu}\cdot s_1+p^{\mu}p^{\nu}\cdot s_2+\\
                    &+\hat{p}p^{\mu}p^{\nu}\cdot s_3+\hat{p}g^{\mu\nu}\cdot s_4+p^{\mu}\gamma^{\nu}\cdot s_5+
                       \gamma^{\mu}p^{\nu}\cdot s_6+\\
                    &+\sigma^{\mu\nu}\cdot s_7+\sigma^{\mu\lambda}p^{\lambda}p^{\nu}\cdot s_8+
                       \sigma^{\nu\lambda}p^{\lambda}p^{\mu}\cdot s_9+
                       \gamma^{\lambda}\gamma^{5}\imath\varepsilon^{\lambda\mu\nu\rho}p^{\rho}\cdot s_{10}.
      \end{split}
      \end{equation}
Here $S^{\mu\nu}$ is an arbitrary spin-tensor depending on the momentum $p$,
$s_{i}(p^2)$ are the Lorentz invariant coefficients, and
$\sigma^{\mu\nu}=\frac{1}{2}[\gamma^{\mu},\gamma^{\nu}]$.
Altogether there are ten independent components in the decomposition
of $S^{\mu\nu}(p)$.

It is known that the $\gamma$-matrix decomposition is complete,
the coefficients $s_{i}$ are free of kinematical
singularities and constraints, and their calculation is rather
simple. However this basis is inconvenient at multiplication and
reversing of the spin-tensor $S^{\mu\nu}(p)$ because the basis
elements are not orthogonal to each other. As a result the
reversing of the spin-tensor $S^{\mu\nu}(p)$ leads to a system of 10
equations for the coefficients.

\item There is another basis used  in  consideration
of the dressed propagator \cite{Kor,AKS,Pas95} $G^{\mu\nu}$. It is
constructed from the following set of operators\footnote{We changed here the
normalization of $\mathcal{P}^{1/2}_{21}$,
$\mathcal{P}^{1/2}_{12}$ for convenience.}
\cite{BDM,Pas95,PvanNie}
\begin{flalign}
      (\mathcal{P}^{3/2})^{\mu\nu}=&g^{\mu\nu}-\frac{2}{3}\frac{p^{\mu}p^{\nu}}{p^2}-
                                     \frac{1}{3}\gamma^{\mu}\gamma^{\nu}+\frac{1}{3p^2}(\gamma^{\mu}p^{\nu}-
                                     \gamma^{\nu}p^{\mu})\hat{p},\notag\\
      (\mathcal{P}^{1/2}_{11})^{\mu\nu}=&\frac{1}{3}\gamma^{\mu}\gamma^{\nu}-
                                          \frac{1}{3}\frac{p^{\mu}p^{\nu}}{p^2}-\frac{1}{3p^2}(\gamma^{\mu}p^{\nu}-
                                          \gamma^{\nu}p^{\mu})\hat{p},\notag\\
      (\mathcal{P}^{1/2}_{22})^{\mu\nu}=&\frac{p^{\mu}p^{\nu}}{p^2},\notag\\
      (\mathcal{P}^{1/2}_{21})^{\mu\nu}=&\sqrt{\frac{3}{p^2}}\cdot\frac{1}{3p^2}(-p^{\mu}+\gamma^{\mu}\hat{p})
                                          \hat{p}p^{\nu},\notag\\
      (\mathcal{P}^{1/2}_{12})^{\mu\nu}=&\sqrt{\frac{3}{p^2}}\cdot\frac{1}{3p^2}p^{\mu}(-p^{\nu}+\gamma^{\nu}\hat{p})
                                          \hat{p}.\label{present}
\end{flalign}
Here
$\mathcal{P}^{3/2}$,$\mathcal{P}^{1/2}_{11}$,$\mathcal{P}^{1/2}_{22}$
are the projection operators while
$\mathcal{P}^{1/2}_{21}$,$\mathcal{P}^{1/2}_{12}$ are nilpotent.
As for their physical meaning, it is clear that
$\mathcal{P}^{3/2}$ corresponds to spin 3/2. The remaining
operators should describe two spin 1/2 representations and
transitions between them.

Let us rewrite the operators
\eqref{present} to make their properties more obvious:
      \begin{flalign}
      (\mathcal{P}^{3/2})^{\mu\nu}=&g^{\mu\nu}-(\mathcal{P}^{1/2}_{11})^{\mu\nu}-(\mathcal{P}^{1/2}_{22})^{\mu\nu},
                                       \notag\\
      (\mathcal{P}^{1/2}_{11})^{\mu\nu}=&3\pi^{\mu}\pi^{\nu},\notag\\
      (\mathcal{P}^{1/2}_{22})^{\mu\nu}=&\frac{p^{\mu}p^{\nu}}{p^2},\notag\\
      (\mathcal{P}^{1/2}_{21})^{\mu\nu}=&\sqrt{\frac{3}{p^2}}\cdot\pi^{\mu}p^{\nu},\notag\\
      (\mathcal{P}^{1/2}_{12})^{\mu\nu}=&\sqrt{\frac{3}{p^2}}\cdot
      p^{\mu}\pi^{\nu}.
      \end{flalign}
Here we introduced the vector
      \begin{equation}
      \pi^{\mu}=\frac{1}{3p^2}(-p^{\mu}+\gamma^{\mu}\hat{p})\hat{p}
      \end{equation}
with the following properties:
      \begin{equation}
      (\pi p)=0,\quad (\gamma\pi)=(\pi\gamma)=1,\quad (\pi\pi)=\frac{1}{3},\quad \hat{p}\pi^{\mu}=-\pi^{\mu}\hat{p}.
      \end{equation}
The set of operators \eqref{present} can be used to decompose the
considered spin-tensor as following \cite{Kor,AKS}:
\begin{eqnarray}
S^{\mu\nu}(p)&=&(S_{1}+S_{2}\hat{p})(\mathcal{P}^{3/2})^{\mu\nu}+
(S_{3}+S_{4}\hat{p})(\mathcal{P}^{1/2}_{11})^{\mu\nu}+
(S_{5}+S_{6}\hat{p})(\mathcal{P}^{1/2}_{22})^{\mu\nu}+ \nonumber
\\
&&(S_{7}+S_{8}\hat{p})(\mathcal{P}^{1/2}_{21})^{\mu\nu}+
(S_{9}+S_{10}\hat{p})(\mathcal{P}^{1/2}_{12})^{\mu\nu}.
\label{p-expan}
\end{eqnarray}
Let us call this basis as $\hat{p}$-basis. It is more convenient
at multiplication since the spin $3/2$ components
$\mathcal{P}^{3/2}$ have been separated from spin $1/2$ ones.
However, the spin $1/2$ components as before are not orthogonal
between themselves and we come to a system of 8 equations when
inverting the \eqref{inverD}. Another feature of decomposition
\eqref{p-expan} is existence of the poles $1/p^2$ in different
terms. So to avoid this unphysical singularity, we should impose
some constraints on the coefficients at zero point.
    \item Let us construct the basis which is the most convenient
at multiplication of spin-tensors. This basis is built from the
operators \eqref{present} and the projection operators
$\Lambda^{\pm}$
      \begin{equation*}
      \Lambda^{\pm}=\frac{\sqrt{p^2}\pm\hat{p}}{2\sqrt{p^2}},
      \end{equation*}
where we assume $p^2>0$.
Ten elements of this basis look as
      \begin{flalign}
      \mathcal{P}_{1}=&\Lambda^{+}\mathcal{P}^{3/2},\,&
      \mathcal{P}_{3}=&\Lambda^{+}\mathcal{P}^{1/2}_{11},\,&
      \mathcal{P}_{5}=&\Lambda^{+}\mathcal{P}^{1/2}_{22},\,&
      \mathcal{P}_{7}=&\Lambda^{+}\mathcal{P}^{1/2}_{21},\,&
      \mathcal{P}_{9}=&\Lambda^{+}\mathcal{P}^{1/2}_{12},\notag\\
      \mathcal{P}_{2}=&\Lambda^{-}\mathcal{P}^{3/2},\,&
      \mathcal{P}_{4}=&\Lambda^{-}\mathcal{P}^{1/2}_{11},\,&
      \mathcal{P}_{6}=&\Lambda^{-}\mathcal{P}^{1/2}_{22},\,&
      \mathcal{P}_{8}=&\Lambda^{-}\mathcal{P}^{1/2}_{21},\,&
      \mathcal{P}_{10}=&\Lambda^{-}\mathcal{P}^{1/2}_{12},
      \label{L-basis}
      \end{flalign}
where tensor indices are omitted. We will call (\ref{L-basis}) as
the $\Lambda$-basis.

Decomposition of a spin-tensor in this basis has the following form:
      \begin{equation}
      S^{\mu\nu}(p)=\sum_{i=1}^{10}\mathcal{P}^{\mu\nu}_{i}\overline{S}_{i}(p^2).\label{l-expan}
      \end{equation}
      
The coefficients $\overline{S}_{i}$ are calculated in analogy with $\gamma$-matrix
decomposition. Besides, we found (with computer analytical computation) the matrix
relating the $\Lambda$-basis with the $\gamma$-matrix basis and convinced
ourselves that this matrix is not singular. So the elements of this basis \eqref{L-basis}
are independent.
It is easy to connect the expansion coefficients \eqref{p-expan} and \eqref{l-expan}
between themselves.
      \begin{flalign}
      \overline{S}_{1}=&S_1+\sqrt{p^2}S_2,\notag\\
      \overline{S}_{2}=&S_1-\sqrt{p^2}S_2,\quad\text{\textit{etc.}}
      \end{flalign}

The $\Lambda$-basis has very simple multiplicative properties
which are represented in the Table~\ref{tt}.
      \begin{table}[ht]
      \caption{Properties of the $\Lambda$-basis at multiplication.\label{tt}}{%
      \begin{tabular}{c|cccccccccc}
      \qquad           & $\mathcal{P}_1$ & $\mathcal{P}_2$ & $\mathcal{P}_3$ & $\mathcal{P}_4$ & $\mathcal{P}_5$ & $\mathcal{P}_6$ & $\mathcal{P}_7$ & $\mathcal{P}_8$ & $\mathcal{P}_9$ & $\mathcal{P}_{10}$\\
      \hline
      $\mathcal{P}_1$    & $\mathcal{P}_1$ & 0&0&0&0&0&0&0&0&0\\
      $\mathcal{P}_2$    &0&$\mathcal{P}_2$&0&0&0&0&0&0&0&0\\
      $\mathcal{P}_3$    &0&0&$\mathcal{P}_3$&0&0&0&$\mathcal{P}_7$&0&0&0\\
      $\mathcal{P}_4$    &0&0&0&$\mathcal{P}_4$&0&0&0&$\mathcal{P}_8$&0&0\\
      $\mathcal{P}_5$    &0&0&0&0&$\mathcal{P}_5$&0&0&0&$\mathcal{P}_9$&0\\
      $\mathcal{P}_6$    &0&0&0&0&0&$\mathcal{P}_6$&0&0&0&$\mathcal{P}_{10}$\\
      $\mathcal{P}_7$    &0&0&0&0&0&$\mathcal{P}_7$&0&0&0&$\mathcal{P}_3$\\
      $\mathcal{P}_8$    &0&0&0&0&$\mathcal{P}_8$&0&0&0&$\mathcal{P}_4$&0\\
      $\mathcal{P}_9$    &0&0&0&$\mathcal{P}_9$&0&0&0&$\mathcal{P}_5$&0&0\\
      $\mathcal{P}_{10}$ &0&0&$\mathcal{P}_{10}$&0&0&0&$\mathcal{P}_6$&0&0&0\\
      \end{tabular}}
      \end{table}
The first six basis elements are projection operators, while the
remaining four elements are nilpotent. We are convinced by direct
calculations that there are no other projection operators besides
indicated.
\end{enumerate}

Now we can return to the Dyson-Schwinger equation \eqref{inverD}.
Let us denote the inverse dressed propagator $(G^{-1})^{\mu\nu}$
and free one $(G^{-1}_{0})^{\mu\nu}$ by $S^{\mu\nu}$ and $S^{\mu\nu}_{0}$
respectively.
Decomposing the $S^{\mu\nu}$, $S_{0}^{\mu\nu}$ and $J^{\mu\nu}$ in $\Lambda$-basis
according to \eqref{l-expan} we reduce the equation \eqref{inverD} to set of
equations for the scalar coefficients
\begin{equation*}
\overline{S}_{i}(p^2)=\overline{S}_{0i}(p^2)+\overline{J}_{i}(p^2)
\end{equation*}
So the values $\overline{S}_{i}$ are defined by the bare propagator and the 
self-energy and may be considered as known.

The dressed propagator also can be found in this form
\begin{equation}
G^{\mu\nu}=\sum_{i=1}^{10}\mathcal{P}^{\mu\nu}_{i}\cdot\overline{G}_{i}(p^2)
\end{equation}
The existing $6$ projection operators take part in the decomposition of $g^{\mu\nu}$:
\begin{equation}
g^{\mu\nu}=\sum_{i=1}^{6}\mathcal{P}^{\mu\nu}_{i}.
\end{equation}
Now solving the equation
\begin{equation*}
G^{\mu\nu}S^{\nu\lambda}=g^{\mu\lambda}
\end{equation*}
in $\Lambda$-basis, we obtain a set of equations for the scalar coefficients
$\overline{G}_{i}$. The equations are easy to solve due to simple
properties of the $\Lambda$-basis --- see Table~\ref{tt}.

The solution is:
\begin{flalign}
\overline{G}_1&=\frac{1}{\overline{S}_1},\quad \overline{G}_2=\frac{1}{\overline{S}_2},\notag\\
\overline{G}_3&=\frac{\overline{S}_6}{\Delta_1},\quad
\overline{G}_4=\frac{\overline{S}_5}{\Delta_2},\quad
\overline{G}_5=\frac{\overline{S}_4}{\Delta_2},\quad
\overline{G}_6=\frac{\overline{S}_3}{\Delta_1},\notag\\
\overline{G}_7&=\frac{-\overline{S}_7}{\Delta_1},\quad
\overline{G}_8=\frac{-\overline{S}_8}{\Delta_2},\quad
\overline{G}_9=\frac{-\overline{S}_9}{\Delta_2},\quad
\overline{G}_{10}=\frac{-\overline{S}_{10}}{\Delta_1},\label{solve}
\end{flalign}
where
\begin{equation}
\Delta_1=\overline{S}_{3}\overline{S}_{6}-\overline{S}_{7}\overline{S}_{10},\qquad
\Delta_2=\overline{S}_{4}\overline{S}_{5}-\overline{S}_{8}\overline{S}_{9}.
\end{equation}
The $\overline{G}_{1}$, $\overline{G}_{2}$ terms which describe
the spin 3/2 have the usual resonance form and could be obtained
from (\ref{p-expan}) as well. As for $\overline{G}_{3} -
\overline{G}_{10}$ coefficients which describe the spin 1/2
contributions, they have a more complicated structure. Let us
consider the denominators of \eqref{solve} in more details.
\begin{flalign}
\Delta_1=&\overline{S}_{3}\overline{S}_{6}-\overline{S}_{7}\overline{S}_{10}=(S_3+\sqrt{p^2}S_4)(S_5-\sqrt{p^2}S_6)-
         (S_7+\sqrt{p^2}S_8)(S_9-\sqrt{p^2}S_{10}),\notag\\
\Delta_2=&\Delta_1(\sqrt{p^2}\to-\sqrt{p^2}).\label{denoms}
\end{flalign}
The appearance of $\sqrt{p^2}$ factor is typical for fermions.
To see it one can find the dressed propagator of the spin 1/2
Dirac particle with a help of the projection operators $\Lambda^{\pm}$.
\begin{equation}
\frac{1}{\hat{p}-m}\Rightarrow \frac{1}{\hat{p}-m-\Sigma(p)}=\Lambda^{+}G^{+}+\Lambda^{-}G^{-}\label{spin1/2}
\end{equation}
The dressed unrenormalizated propagator is:
\begin{flalign}
(G^{+})^{-1}=&-m-A(p^2)+\sqrt{p^2}(1-B(p^2)),\notag\\
(G^{-})^{-1}=&-m-A(p^2)-\sqrt{p^2}(1-B(p^2)),\label{dirac}
\end{flalign}
where $A,$ $B$ are the self-energy components:
\begin{equation*}
\Sigma(p)=A(p^2)+\hat{p}B(p^2).
\end{equation*}
Note that use of the projection operators $\Lambda^{\pm}$ in the
case of Dirac particle  is useful but not necessary technical
step. After renormalization thus obtained propagator will coincide
with expression given in any textbook.
As for the apparent branch point $\sqrt{p^2}$, it is canceled in total
expression \eqref{spin1/2}. The same is true for the dressed Rarita-Schwinger 
propagator \eqref{solve}.

Note that the propagator's denominators \eqref{denoms} are not
similar by their structure to the Dirac case \eqref{dirac}. The
nearest analogy for the Rarita-Schwinger field propagator is the
joint dressing of two Dirac fermions with account for their mutual
transition. Probably this analogy will be useful in the renormalization.

\section*{Conclusion}

Thus we obtained the simple analytical expression \eqref{solve}
for the interacting Rarita-Schwinger field propagator which
accounts for all spin components. To derive it we introduced
the spin-tensor basis \eqref{L-basis} with very simple
multiplicative properties. This basis is singular and it seems
unavoidable (recall the vector field case). Nevertheless the
singularity of a basis is not so big obstacle in its use.
We did not suppose here any symmetry properties of the self-energy
$J^{\mu\nu}$ restricting ourselves by general case. Of course the 
concrete form of interaction will lead to some symmetry of $J^{\mu\nu}$
and it will be important at renormalization.

Note some features of the obtained answer \eqref{solve}. First,
if there is only one spin 1/2 pole it can not appear in
$\Delta_{1}$ and $\Delta_{2}$ simultaneously. Second, it is well
known that $\Psi^{\mu}$ field contains two spin 1/2 components
(two irreducible representations) associated with
$\mathcal{P}^{1/2}_{11}$ and $\mathcal{P}^{1/2}_{22}$ operators.
However it turns out that these two representations are not completely
independent since they have common denominators
$\Delta_{1}$, $\Delta_{2}$.

The obtained dressed propagator \eqref{solve} solves an algebraic
part of the problem, the following step is renormalization. Note that
the investigation of dressed propagator is the alternative for more
conventional method based on equations of motion (see, \textit{i.e}
Ref.~\cite{Pas99} and references therein).
The natural requirement for the renormalization is that the spin
$1/2$ components should remain unphysical after dressing. In other
words the denominators $\Delta_1$, $\Delta_2$ should not acquire
of zero in the complex energy plane. However a problem of
the renormalization needs a more careful consideration.

\section*{Acknowledgements}

We are indebted D. V. Naumov for reading the manuscript and useful
remarks. We thank G. Lopez Castro and D. V. Ahluwalia-Khalilova for
comments and references.


\vspace*{6pt}

\begin{thebibliography}{10}
\bibitem{Rar-Sch}
W.~Rarita and J.~Schwinger.
\journal{Phys. Rev.} \textbf{60} (1941) 61.

\bibitem{JohnSud61}
K. Johnson and E.~C.~G. Sudarshan. \journal{Ann. Phys.} (N.Y.)
\textbf{13} (1961) 126

\bibitem{VeZw69}
G.~Velo and D.~Zwanzinger. \journal{Phys. Rev.}
\textbf{186} (1969) 267, 1337

\bibitem{Kor}
C. L. Korpa. \journal{Heavy Ion Phys.}
\textbf{5} (1997) 77.

\bibitem{Pas98}
V.~Pascalutsa. \journal{Phys. Rev.}
\textbf{D58} (1998) 096002.

\bibitem{Pas01}
V. Pascalutsa. \journal{Phys. Lett.}
\textbf{B503} (2001) 86

\bibitem{Cas} G. Lopez Castro and A. Mariano. 
\journal{Phys. Lett.}
\textbf{B517} (2001) 339

\bibitem{Kir} M. Kirchbach and D. V. Ahluwalia. 
\journal{Phys. Lett}
\textbf{B529} (2002) 124

\bibitem{AKS}
A. N. Almaliev, I. V. Kopytin and M. A. Shehalev.
\journal{J. Phys. G.}
\textbf{28} (2002) 233

\bibitem{AU}
A.~Aurilia, H.~Umezava.
\journal{Phys. Rev.} \textbf{182} (1969) 1686


\bibitem{NEK}
L.~M.~Nath,B.~Etemadi and J.~D.~Kimel.
\journal{Phys. Rev.} \textbf{D3} (1971) 2153

\bibitem{BDM}
M.~Benmerrouche, R.~M.~Davidson and N.~C.~Mukhopadhyay.
\journal{Phys. Rev.} \textbf{C39} (1989) 2339.

\bibitem{Pas95} V. Pascalutsa and O. Scholten. 
\journal{Nuc. Phys.} \textbf{A591} (1995) 658

\bibitem{KS}
S.~Kondratyuk and O.~Scholten.
\journal{Phys. Rev.} \textbf{C62} (2000) 025203

\bibitem{MaRo}
P. Maris and C. D. Roberts.
\journal{Int. J. Mod. Phys.}
\textbf{E12} (2003) 297

\bibitem{PvanNie}
P.~van~Nieuwenhuizen.
\journal{Phys. Rep.} \textbf{68} (1981) 189

\bibitem{Pas99} V. Pascalutsa and Rob Timmermans.
\journal{Phys.Rev.}
\textbf{C60} (1999) 042201

\end{thebibliography}
\end{document}